# Mechanism of CO-oxidation on Pd/CeO$_2$(100): The unique surface-structure of CeO$_2$(100) and the role of peroxide

Yongeseon Kim[1], Hosik Lee[2,*], and Ja Hun Kwak[1,*]

[1]School of Energy and Chemical Engineering, Ulsan National Institute of Science and Technology (UNIST), 50 UNIST-gil, Ulsan 44919, Republic of Korea

[2]School of Mechanical, Aerospace and Nuclear Engineering, Ulsan National Institute of Science and Technology (UNIST), 50 UNIST-gil, Ulsan 44919, Republic of Korea

[*]*Corresponding authors:* hslee@unist.ac.kr, jhkwak@unist.ac.kr

**Abstract**

Understanding the atomic mechanism of low-temperature CO oxidation on a heterogeneous catalyst is challenging. We performed density functional theory (DFT) calculations to identify the surface structure and reaction mechanism responsible for low-temperature CO oxidation on Pd/CeO$_2$ (100) surfaces. DFT calculations reveal the formation of a unique zigzag chain structure by the oxygen and Ce atoms of the topmost surface of CeO$_2$(100) with Pd atoms located between the zigzag chains. O$_2$ adsorbed on such Pd atoms is stable in the presence of CO but plays a very important role in lowering the activation barrier for low-temperature CO oxidation by forming a square-planar PdO$_4$ structure and facilitating further O$_2$ adsorption. *In-situ* Raman spectroscopy studies confirm the adsorbed oxygen species to be peroxides. The calculated activation barrier for CO oxidation, based on the mechanism suggested by these unique structures and peroxides, is 31.2 kJ/mol, in excellent agreement with our experimental results.

Ceria ($CeO_2$) plays an important role as an active support of heterogeneous catalysts in many applications, owing to its excellent redox properties and oxygen storage capacity[1-3]. For example, it is used in automotive exhaust treatment (three-way catalyst)[4-6], water-gas shift reaction[7,8], and steam reforming of alcohols[2,9]. Pd on $CeO_2$ catalyzes low-temperature CO oxidation as well as NO reduction[2,10]. However, the active sites in the structure have been a subject of dispute[11-14]. In general, the low-temperature CO oxidation is speculated to follow the Mars-van Krevelen mechanism, where CO directly reacts with the surface oxygen ($O_s$), thus forming $CO_2$ and leaving behind surface oxygen vacancies ($O_v$), which are subsequently filled by gas-phase $O_2$[15,16]. Since the surface oxygen vacancies should first be occupied by molecular oxygen, a surface-attached superoxide or peroxide is commonly observed[17-20]. The entire reaction sequence is very complex owing to the presence of surface CO, $CO_2$, carbonate, superoxide, and peroxide species. These species have been experimentally observed by various spectroscopic techniques such as Fourier-transform infrared (FT-IR)[17], Raman[19-21], and polarization-dependent infrared reflection-absorption spectroscopy (IRRAS)[18]; however, unambiguous experimental identification of the active site is very difficult because of the complexity and dynamic aspects of the low-temperature CO oxidation reactions on Pd/$CeO_2$.

Theoretical approaches based on density functional theory (DFT) can be used to understand the reaction dynamics at the atomic level, because they allow one to simulate how the molecules interact with a specific surface structure and thus unravel the details of low-temperature CO oxidation. Although low index facets such as (100), (110), and (111) have been studied theoretically and experimentally[16,18,19,22-32], the origin of the excellent low-temperature CO oxidation on the (100) facet is still not clear in mechanistic aspects. In this work, we performed first-principles DFT calculations to understand the mechanism of low-temperature CO oxidation on Pd/$CeO_2$. Among the low-index surfaces, we found that the stoichiometric $CeO_2$ (100) surface has vacancies at surface oxygen ($O_s$) sites that decrease the surface energy.

Half of the $O_s$ sites are vacant, whereas the remaining ones form a zigzagged chain with the Ce atoms. In the valleys between the zigzag chains, a Pd atom can be placed and thus a Pd-O-Ce structure is established. The activation barrier for CO oxidation on this structure is comparable to that determined experimentally. Further, we also discuss the formation of peroxide and its role in CO oxidation. To experimentally demonstrate the theoretical results, we prepared 0.1% Pd/CeO$_2$ catalysts from cubic and octahedral CeO$_2$ nanoparticles with dominant (100) and (111) facets, respectively. The observed activation barrier for CO oxidation is consistent with the calculated one and those reported previously. *In-situ* Raman spectroscopic analysis supports the theoretical findings on the role of peroxide during the low-temperature CO oxidation. Our combined theoretical and experimental study provides a good picture of low-temperature CO oxidation on Pd/CeO$_2$ with details on the associated reactions.

**Results and Discussions**

**DFT computations.** Among the low-index (100), (111), and (110) facets, we found that the CeO$_2$ (100) surface can have possible active sites for CO oxidation owing to its surface normal dipole. Therefore, we focused on the CeO$_2$ (100) surface as the active surface for CO oxidation, in accordance with previous experiments and calculations[16, 33-35]. Tasker categorized the surfaces of ionic crystals into three types, as shown in Fig. 1 a[36]. Type 1 stacking sequences such as in the CeO$_2$ (110) surface have the same number of cations and anions in each plane and type 2 stacking sequences such as that of the CeO$_2$ (111) surface show a symmetric distribution of the cationic and anionic layers leading to no net surface-normal polarization. However, the surface-normal dipole remains in type 3 stacking sequences such as that of the CeO$_2$ (100) surface, and hence this type of surface will be forbidden. Tasker insisted that the polar surface, here, the stoichiometric CeO$_2$ (100) surface, undergoes charge redistribution due to vacancies or adatoms on the opposite surface[36]. In other words, only half of surface oxygen atoms should be located in one (100) surface and the other surface oxygen atoms in opposite (100) surface according to Tasker's suggestion. The surface is different from that in literature in that all the surface oxygen atoms are located in one (100) surface and surface Ce atoms are exposed in opposite surface. We performed DFT calculations according to this idea. At first, we calculated the energies of the 1×1 (100) surfaces with four different configurations, which are shown in Fig. 1 b. Red, gray, and green spheres represent oxygen, cerium, and the topmost surface oxygen atoms, respectively. Four topmost surface atoms appear on the 1×1 (100) surface. The unseen green atoms in Fig. 1 b are moved to the opposite side to sustain the stoichiometry with the same number of atoms in our unit cell. Obviously, the most polarized surface has the highest energy and the most depolarized surface has the lowest energy (Fig. 1 b). We adopted a surface structure with the lowest energy as the model for the (100) surface in the subsequent calculations. We then expanded our unit cell to a 2×2 structure. The most stable structure of the seven configurations considered is shown in Fig. 1 c (left). (For the detailed structure of the configurations, see Fig. S1.) The topmost Ce and O atoms form a zigzag chain structure, which is represented by black and green spheres in the model. Valleys appear between the zigzag chain structures owing to missing surface O atoms, and these are moved to the opposite side for charge redistribution.

Then, we carried out calculations on possible Pd positions in the valleys (Fig. S2), and the most stable Pd-O-Ce structures are shown in Fig. 1 c (center panel). Here, we focus on single atom Pd because of sufficiently low surface Pd density of our catalysts (0.35 and 0.71 Pd-atom/nm$^2$ for 0.1% Pd/CeO$_2$(100) and 0.1% Pd/CeO$_2$(111), respectively.). Surprisingly, when Pd is positioned in the valleys, the calculated formation energy of the structure is 352 kJ/mol, which implies that the valleys are very favorable sites for locating Pd. Note that the results of our calculations indicate that the stoichiometric $CeO_2$ (100) surface naturally has vacant sites in the form of valleys between zigzagged $Ce-O_s$ chains, which are not available on (110) and (111) surfaces without O defects, and Pd atoms can reside in these valleys.

With this model, we found that the computed activation energy for CO oxidation on our (100) surface is within the range of characteristic activation energy of 30–57 kJ/mol reported in previous studies[11, 12, 37, 38]. First, we calculated the CO adsorption barrier for four different surfaces, viz., $CeO_2$ (100), $Pd/CeO_2$ (100), $CeO_2$ (111), and $Pd/CeO_2$ (111) (computational details may be found in the Supporting Information). Except for the $Pd/CeO_2$ (100) surface with a Pd-Ce-O structure, the surfaces have a higher CO adsorption barrier than the experimental CO activation barrier (Figure 1 d). The computed CO adsorption barrier for the $Pd/CeO_2$ (100) surface is 31.2 kJ/mol. Although this value is slightly lower than the previously reported experimental values, it is close to our experimental value of 34–35 kJ/mol. The results of the calculation are shown as vertical red bars and those of the previous experiments are indicated by a horizontal blue bar in Fig. 1 d; our experimental observation is marked with green diamond on the same plot.

The CO adsorption barrier on the $Pd/CeO_2$ (100) surface depends on the reaction path. For the Pd-O-Ce structure shown in the central panel of Fig. 1 c, the CO adsorption barrier is 7.7 kJ/mol and surface-attached $CO_2$ is formed with one $O_s$ following Mars-van Krevelen mechanism. However, the barrier for the detachment of $CO_2$ is 61.8 kJ/mol. This $CO_2$ detachment barrier is significantly higher than the experimentally determined activation energy (34–35 kJ/mol) and therefore, this reaction pathway is not plausible. Thus, we considered another pathway in which the $O_2$ molecule is adsorbed before CO adsorption, which seems to be a reasonable possibility. In this case, the adsorption barrier of the $O_2$ molecule is 0 kJ/mol and the adsorbed $O_2$ turns into surface peroxide, as shown on the right-hand side of Fig. 1 d. With this peroxide, the Pd atom establishes a square-planar $PdO_4$ structure with bonding between Pd, a subsurface oxygen, and peroxide (Fig. 1 d). It is noteworthy that the square-planar $PdO_4$ and peroxide resist CO. Any CO molecule approaching the square-planar $PdO_4$ and peroxide is expelled from the surface without adsorption. This indicates that the $PdO_4$ structure and nearby peroxide are not directly involved in the reaction. Further calculations for the reaction show that CO reacts with $O_s$ to form surface carbonate and the energy of detachment of $CO_2$ from the carbonate is greater than 280 kJ/mol. This large $CO_2$ detachment energy is consistent with a stable carbonate species that has been observed experimentally[11].

To find a lower overall CO oxidation barrier, we constructed another peroxide. The peroxide-attached $Pd/CeO_2$ (100) surface readily accepts an $O_2$ molecule with an adsorption barrier of 0 kJ/mol. Located at the corner of the zigzag valley, this peroxide hinders the formation of a stable carbonate by repelling the newly formed $CO_2$ molecule; we refer to this peroxide as the repelling peroxide. A detailed mechanism for the low-temperature CO oxidation on a $Pd/CeO_2$ (100) surface with the repelling peroxide is shown in Fig. 2. The first CO molecule approaches oxygen at the reaction site. The corresponding CO adsorption barrier is 31.2 kJ/mol, which is the highest barrier during the CO oxidation process. We identify the structure of CO adsorption barrier 31.2 kJ/mol as a transition state during a $CO_2$ activation process. In the process, initial state is a state where a CO molecule apart from the $Pd/CeO_2$ (100) surface and the final state is a state that a $CO_2$ molecule is formed and depart from the surface. The process is exothermic and the energy difference between initial and final state is 348.9 kJ/mol. When the distance between the CO molecule and the surface is less than the transition structure (blue dash line of Fig. S4 (b)), the $CO_2$ molecule is formed with $O_s$ spontaneously following Mars-van Krevelen mechanism. Otherwise, the CO molecule shows physisorption with binding energy of 13 kJ/mol. For reaction energy diagram of the 1$^{st}$ and 2$^{nd}$ CO2 molecules, see Fig. S5. Note that the CO adsorption barrier without the repelling peroxide is 7.7 kJ/mol whereas the energy for the detachment of $CO_2$ from the formed carbonate is 280 kJ/mol. However, with the repelling peroxide, the generated $CO_2$ can depart from the surface without a barrier, owing to the repulsive force of the repelling peroxide. Thus, the repelling peroxide heightens the CO adsorption barrier, while lowering the $CO_2$ detachment barrier significantly. The second CO molecule approaches the other reaction site with an adsorption barrier of 14.5 kJ/mol and reacts with $O_s$. The generated second $CO_2$ molecule departs from the surface without a barrier, leaving behind an oxygen vacancy. Then, there are two oxygen vacancies and the $O_2$ molecule can move to the vacancy site without a barrier, as shown in Fig. 2. This adsorbed oxygen is a reactive peroxide that dissociates readily, in contrast to the peroxide of the $PdO_4$ structure and the repelling peroxide. The dissociation barrier of the reactive peroxide is 5.53 kJ/mol and the reaction cycle is closed.

Before moving on to the description of the experimental results, we would like to emphasize our theoretical findings, which can be compared with the experimental ones. Our DFT results suggest the assisting role of the peroxide, which does not react directly with CO but facilitates low-temperature CO oxidation by facilitating easy desorption of $CO_2$. The low-temperature CO oxidation occurs on the $Pd/CeO_2$ (100) surface and the overall activation barrier is calculated to be ~31.2 kJ/mol, which is close to the experimentally determined value, which will be discussed later in detail. To verify the DFT results, we prepared $Pd/CeO_2(100)$ and $Pd/CeO_2(111)$, by impregnating Pd into pre-prepared $CeO_2(100)$ and $CeO_2(111)$. Further, we conducted *in-situ* Raman spectroscopy to elucidate the role of peroxide during the CO oxidation. In general, adsorbed oxygen species, such as peroxides or superoxides, are simply considered to be reactive species. Moreover, there is considerable debate as to which of the species is reactive. Guzman et al.[39] and Lohrenscheit et al.[40] showed that the Raman band of the peroxide at the isolated defect site (at 830 cm$^{-1}$) does not undergo significant change under the flow of CO and suggested that this peroxide (at 830 cm$^{-1}$) on Au/CeO$_2$ is not a reactive species. Instead, superoxide or other peroxide species were considered as reactive species. On the contrary, Lee et al.[41] reported a strong peroxide peak at 830 cm$^{-1}$ in the Raman spectrum when the reaction of CO and O$_2$ was active on Au/CeO$_2$, and they suggested a reaction between the peroxide and CO. The amount of labile oxygen atoms and the catalytic activity of the 0.1% Pd/CeO$_2$ catalysts were further compared via CO-temperature-programmed reduction (CO-TPR) and temperature-programmed and steady-state CO oxidation (Fig. 3).

**CO-TPR on 0.1% Pd/CeO$_2$ catalysts.** The CO-TPR was carried out to compare the amount of labile oxygen atoms of 0.1% Pd/CeO$_2$ catalysts and results are shown in Fig. 3 a. The CO-TPR results support the hypothesis that the reaction occurs on the 0.1% Pd/CeO$_2$ (100) surface. 0.1% Pd/CeO$_2$(100) shows reduction peaks at 93, 159, and 290 °C. In case of 0.1% Pd/CeO$_2$(111), reduction peaks appeared at 87 and 379 °C. Both the samples show similar peak positions for the low-temperature reduction with the same onset temperature; however, a significantly higher amount of CO was consumed by 0.1% Pd/CeO$_2$(100) than by 0.1% Pd/CeO$_2$(111). In case of 0.1% Pd/CeO$_2$(100), the amount of CO consumption below 200 °C is 52.5 µmol/g$_{cat}$. Because all the Pd initially exists as PdO before CO-TPR (Fig. S12), the amount of CO consumption, required to reduce PdO to Pd, should be 9.4 µmol/g$_{cat}$. Therefore, the amount of excess CO consumption (43.1 µmol/g$_{cat}$) is due to oxygen from CeO$_2$(100). By contrast, 0.1% Pd/CeO$_2$(111) showed similar CO consumption (13.8 µmol/g$_{cat}$) as that expected for actual Pd loading, implying that the CeO$_2$(111) surface has a lower amount of reactive oxygen atoms under CO-TPR. Note that the lowest CO-TPR peak positions of 0.1% Pd/CeO$_2$(100) and 0.1% Pd/CeO$_2$(111) are similar despite their significant difference in peak intensity. This implies that 0.1% Pd/CeO$_2$(111) has the same reactive sites as 0.1% Pd/CeO$_2$(100) but in a much smaller number. The intense peak at 379 °C on 0.1% Pd/CeO$_2$(111) can be related to bulk lattice oxygen.[11] Considering CO-TPR result of CeO$_2$ supports (Fig. S13), lattice oxygen activation by Pd is more striking on $CeO_2(100)$ than $CeO_2(111)$.

**CO oxidation reactivity.** The catalytic CO oxidation activities of the two catalysts were compared via temperature-programmed CO oxidation experiments (Fig. 3 b). CO oxidation starts at ~50 °C on both 0.1% $Pd/CeO_2(100)$ and 0.1% $Pd/CeO_2(111)$ catalysts. Although the conversion of CO starts at a similar temperature, the conversion on 0.1% $Pd/CeO_2(100)$ is significantly higher than on 0.1% $Pd/CeO_2(111)$. For a quantitative comparison of the catalytic activity and activation energy, we estimated the steady-state activity and the results are shown in the inset of Fig. 3 b. Interestingly, both catalysts exhibit practically the same activation energy of 34–35 kJ/mol, while the turnover frequency (TOF) (calculated based on the Pd weight) for the reaction on 0.1% $Pd/CeO_2(100)$ is more than four times higher than that of the reaction on 0.1% $Pd/CeO_2(111)$ (Table 1). It is noteworthy that the experimentally estimated activation barrier is similar to calculated activation barrier (31.2 kJ/mol). This value also falls within a range of previously reported activation energy of 30–57 kJ/mol for CO oxidation on $Pd/CeO_2$[11, 12, 37, 38]. From the CO-TPR and temperature-programmed CO oxidation experiments, we obtained results consistent with those of our calculations, and we conclude that the 0.1% $Pd/CeO_2$ (100) surface is the origin of low-temperature CO oxidation with the corresponding barrier being ~31.2 kJ/mol, as obtained from the calculations. The results of a second calculation that suggested that there are stable peroxide species can be verified by *in-situ* Raman spectroscopy, whereby the surface peroxide can be identified.

*In-situ* **Raman spectroscopy.** *In-situ* Raman spectroscopy was carried out for structural analysis and to observe the behavior of adsorbed oxygen species on reduced 0.1% $Pd/CeO_2$ under the flow of He, CO + $O_2$ and $O_2$ (Fig. 3 c, d). Under the He flow, 0.1% $Pd/CeO_2$ samples show a Raman band at 462 cm$^{-1}$ and weak bands near 260, 598, and 1174 cm$^{-1}$, which are attributed to the $F_{2g}$ mode of the cubic fluorite structure, second-order transverse acoustic (2TA) mode, defect-induced (D) mode, and second-order longitudinal optical (2LO) mode, respectively[19]. However, the band corresponding to Pd-O-Ce was not observed, which we attribute to low Pd loading. In the presence of the reactant gas mixture (1% CO + 2.5% $O_2$ in He), prominent additional Raman bands at 830 and 858 cm$^{-1}$ were detected on the 0.1% Pd/CeO$_2$(100) samples and these bands are assigned to adsorbed peroxide on isolated and clustered two-electron defect sites[19, 20, 39, 41] (inset, Fig. 3 c), whereas, no noticeable change was found on 0.1% Pd/CeO$_2$(111) samples (inset, Fig. 3 d). The significant contrast between the peroxide peaks of 0.1% Pd/CeO$_2$(100) and 0.1% Pd/CeO$_2$(111) indicates the critical role of peroxide in the low-temperature CO oxidation on the 0.1% Pd/CeO$_2$ surface, which could not be deduced from the CO-TPR and temperature-programmed CO oxidation experiments. Although the critical role of the peroxide is not clear when only the Raman results are considered, it can be understood by taking our calculations into account. According to our calculations, two kinds of peroxides are stable upon CO adsorption. One is a peroxide species that forms a PdO$_4$ structure and the other is a repelling peroxide. The repelling peroxide lowers the energy for $CO_2$ desorption to 0 kJ/mol, with the repulsive force counteracting a slight increase in the CO adsorption energy. Our Raman results provide evidence for the existence of peroxides on the highly active 0.1% Pd/CeO$_2$ surface, especially on the (100) surface. The peroxides and superoxide were so far commonly considered to be reactive species[19, 39-41]. However, the prominent peroxide peaks from 0.1% Pd/CeO$_2$(100) in the presence of the reactant gas mixture indicate that they are not reactive species. If they were reactive species, their peak intensity should decrease with the supply of CO; contrarily, the peak intensity increased. The simultaneous appearance of the peroxide peak and the high CO oxidation rate, in accordance with our calculation, strongly support our picture of low-temperature CO oxidation on the 0.1% Pd/CeO$_2$ (100) surface.

In spite of combined and consistent picture from our theoretical and experimental work explained above, it is hard to confirm the picture due to lack of direct experimental evidence on Pd status. Scanning transmission electron microscopy (STEM) and Extended X-ray adsorption fine structure (EXAFS) may show the direct experimental evidence of the unique Pd-O-Ce structure, but not available now and further studies are required. Meanwhile, other reaction pathways with Pd cluster/nanoparticle may contribute to the low temperature CO oxidation reaction somewhat. Our conclusions do not rule out the possibility. However, we stress that the unique Pd-O-Ce structure can contribute to the low temperature CO oxidation reaction at least partly. Moreover, we believe the pathway on the unique Pd-O-Ce structure is main contribution to the low temperature CO oxidation reaction in that our overall experimental and calculational results are consistent and show the very low activation barrier of ~30 kJ/mol. As is known, whole reaction is governed by the lowest activation barrier pathway and we believe we found the pathway.

In summary, we compared two different facets of cerium oxide, (100) and (111), to understand the origin of the superior CO oxidation activity on the (100) facet. Our DFT calculations suggest a unique surface structure of $CeO_2$(100) with the formation of a very stable Pd-O-Ce structure on the surface. Combined with *in-situ* spectroscopic analysis, we demonstrated that the peroxides are detected only on the (100) surface and three types of peroxides are involved in the overall CO oxidation reaction mechanisms, and that they play different roles. Firstly, the peroxide forms a $PdO_4$ structure with Pd and remains inert in the presence of CO. The repelling peroxide facilitates desorption of $CO_2$. Lastly, the reactive peroxide dissociates into two oxygen vacancies and then reacts with CO. Our theoretical and experimental results are in agreement and they provide new insights into the role of peroxide in the reaction mechanism of low-temperature CO oxidation.

## Method

### DFT computations

The density functional theory calculations were performed with the local density functional approximation (LDA) as implemented in VASP (Vienna *ab initio* simulation Package[42]). A Hubbard U term was added to the LDA functional (LDA + U) employing the formalism by Dudarev *et al.*[43], in which only the difference ($U_{eff} = U - J$) between the Coulomb $U$ and exchange $J$ parameter arises. Spin-polarized calculations are performed with a plane-wave basis set with a kinetic energy cut-off of 450 eV. We adopted $U_{eff} = 4.0$ eV for Ce *4f* states. These values are reported to provide appropriate properties such as the lattice parameters of $CeO_2$ and $Ce_2O_3$, electronic structure including the band gap, reaction energy of $2CeO_2 \rightarrow Ce_2O_3 + (1/2)O_2$, and vacancy formation energies[43]. For all the surface calculations, a model with a periodic slab with a (2×2) configuration for the (100) and (110) surfaces and a (3×3) configuration for the (111) surface was used. A Monkhorst-Pack 2×2×1 mesh was used for Brillouin zone integration. The $CeO_2$ (100), (110), (111) slab models were of four, three, and three layers in thickness, respectively, and the vacuum gap was set to 10 A. For determining the CO and $O_2$ adsorption barriers, we performed static calculations in which the molecules were made to approach each type of surface. Optimal CO and $O_2$ positions are shown in the Supplementary information (Fig. S3). The vacuum levels were set to the energy of the system such that they would not vary significantly at certain separation distances between the surface and molecules. The vacuum level was set to zero for the adsorption energy. The molecules started to bond to the surface at a distance lesser than the blue dashed line (Fig. S4). The migration barriers for $O_2$ dissociation were calculated using the nudged elastic band (NEB) method (Fig. S7).

### Preparation of catalysts

CeO$_2$(100) and (111) were synthesized by a hydrothermal process in Teflon-lined stainless steel autoclaves as reported previously[19]. 0.1% Pd/CeO$_2$ catalysts with 0.1 wt% Pd were prepared by conventional incipient wetness impregnation using a solution of Pd(NH$_3$)$_4$(NO$_3$)$_2$. After impregnation, the samples were calcined in air at 673 K for 4 h.

**Characterization of the catalyst**

Scanning electron microscopy (SEM) was performed using an S-4800 (Hitachi) system operating at 5 kV. Transmission electron microscopy (TEM) was performed using a JEOL TEM-2100 system operating at 200 kV. (Fig. S8, 9).

X-ray diffraction (XRD) patterns of the synthesized CeO$_2$ powder were collected on a D8 Advance (Bruker) using Cu $K\alpha$ radiation ($\lambda = 1.5406$ Å) at 40 kV and 40 mA, with a step size of 0.05 in $2\theta$ and a time per step of 0.5 sec in the $2\theta$ range from 10 to 110 ° (Fig. S10). Brunauer-Emmett-Teller (BET) surface area of the CeO$_2$ powder was determined via N$_2$ adsorption experiments using a BELSORP-max system (Bel Japan) (Fig. S11). Prior to N$_2$ adsorption, the samples were treated in a dry N$_2$ flow at 150 °C for 4 h.

X-ray adsorption fine structure (XAFS) at Pd K-edge (24350 eV) were performed in fluorescence mode at the 7D beamline of the Pohang Accelerator Laboratory (PLS-II) (Fig. S12)

Raman analysis was performed on WITec alpha300R Micro-imaging Raman Spectrometer equipped with a 532-nm (3 mW) Nd-YAG excitation laser and 40× Nikon objective (NA = 0.6). A spectrometer with a grating of 1800 gr/nm was used. The spectral acquisition was executed with 20 scans at 3 s/scan using an electrically cooled CCD detector.

*In-situ* Raman spectroscopy was performed in a high-temperature reaction chamber (Harrick scientific). The 0.1% Pd/CeO$_2$ samples were pretreated at 350 °C for 0.5 h under 20% O$_2$/He (60 cc min$^{-1}$) followed by reduction at 350 °C for 0.5 h under 10% H$_2$/He (60 cc min$^{-1}$). The samples were cooled in He (60 cc min$^{-1}$) to room temperature. Reduced samples were exposed to 2.5% O$_2$ + 1% CO/He and 2.5% O$_2$/He (60 cc min$^{-1}$) with an exposure time of 5 min in each case.

For CO-TPR experiments, 0.05 g of the sample was pre-treated at 350 °C for 0.5 h under 20% O$_2$/He flow (60 cc min$^{-1}$). After cooling the sample to room temperature, it was purged with He (60 cc min$^{-1}$) for 0.5 h. The reduction was carried out in 0.2% CO/He (60 cc min$^{-1}$) flow with a heating rate of 10 °C min$^{-1}$. CO$_2$ produced during the catalyst reduction was adsorbed with CaX zeolite. The remaining CO was converted to methane using Ni/Al$_2$O$_3$ catalysts with H$_2$ at 400 °C and then flown directly into the flame ionization detector (FID) of an Agilent 7820A gas chromatograph (GC). The amount of CO consumed was determined from the FID signal intensities, with the FID sensitivity factor calibrated using a 100-μL pulse of 10% CO in He.

H$_2$-temperature programmed reduction (TPR) experiments were performed using a homemade stainless steel furnace block equipped with a cartridge heater (Fig. S14). Typically, 0.05 g of the sample was loaded into a linear quartz tube along with quartz wool. Prior to the experiment, the sample was pre-treated at 350 °C for 0.5 h and then cooled to −20 °C under a flow of 20% O$_2$/He (60 cc min$^{-1}$). After stabilization of the signal of the thermal conductivity detector (TCD) for 2 h, reduction was carried out under a flow of 2% H$_2$/Ar (50 cc min$^{-1}$) at a heating rate of 10 °C min$^{-1}$. The amount of H$_2$ consumed was determined from the TCD signal intensities calibrated using CuO/SiO$_2$ references. A cold trap (dry ice in acetone) was used to remove water produced during the catalyst reduction.

**CO oxidation**

For determining the activity of CO oxidation, samples were reduced as described in previous works[11, 45]. A reduction temperature of 350 °C was selected according to the $H_2$-TPR result. The CO oxidation activity was evaluated via temperature-programmed reaction methods in a fixed bed reactor using 0.1 g of the catalyst after reduction treatment. Before measuring the activity, the catalysts were pretreated at 350 °C for 0.5 h under 20% $O_2$/He (60 cc min$^{-1}$) followed by reduction at 350 °C for 0.5 h under 10% $H_2$/He (60 cc min$^{-1}$). The activity was measured using a feed gas mixture containing 1% CO and 2.5% $O_2$ in He (60 cc min$^{-1}$) by ramping up the temperature at the rate of 5 °C min$^{-1}$. The outlet gases were analyzed using a GC (Agilent 7820A) using a HP-PLOT Q column and TCD. Steady state activity was measured in a quartz flow reactor with 0.02 g of the catalyst supported by quartz wool. To achieve a reasonable conversion, the catalysts were diluted with an inert material (quartz chips) when required. The activation energy was calculated with the specific reaction rate as described in a previous work[45] by averaging the conversions at 20, 40, and 60 min.

**Data availability**

The data that support the findings of this study are available from the corresponding author on request.

## Acknowledgements

We acknowledge the financial support from the National Research Foundation (NRF) (No. 2016R1A5A1009405, 2017R1A2B4007310, and 2017R1D1A1B03028004) and the Ministry of Trade, Industry & Energy (10050509, N0001754). We also thank the HEESUNG CATAYSTS CORP. for their financial support. This work was supported by the Supercomputing Resources of KISTI, including technical support (KSC-2018-C2-0007). We thank Sung June Cho for help in EXAFS data analysis and Eunjung Jang for DRIFT analysis.

## Author contributions

J.H.K. conceived and designed the project. H.S.L. performed DFT computations. Y.S.K. carried out the catalyst preparation, characterizations, and CO oxidation tests. All the authors participated in discussions and writing of the paper.

## Additional Information

**Competing interests**

The authors declare no competing interests.

Table 1. Catalytic activities of 0.1% Pd/CeO$_2$ catalysts in CO oxidation

| catalyst | CO oxidation at 80 °C | |
| --- | --- | --- |
| | TOF ($\times 10^4$ mol g$_{Pd}^{-1}$ s$^{-1)}$) | E$_a$ (kJ/mol) |
| 0.1% Pd/CeO$_2$(100) | 11.72 | 35 |
| 0.1% Pd/CeO$_2$(111) | 2.71 | 34 |

**a**

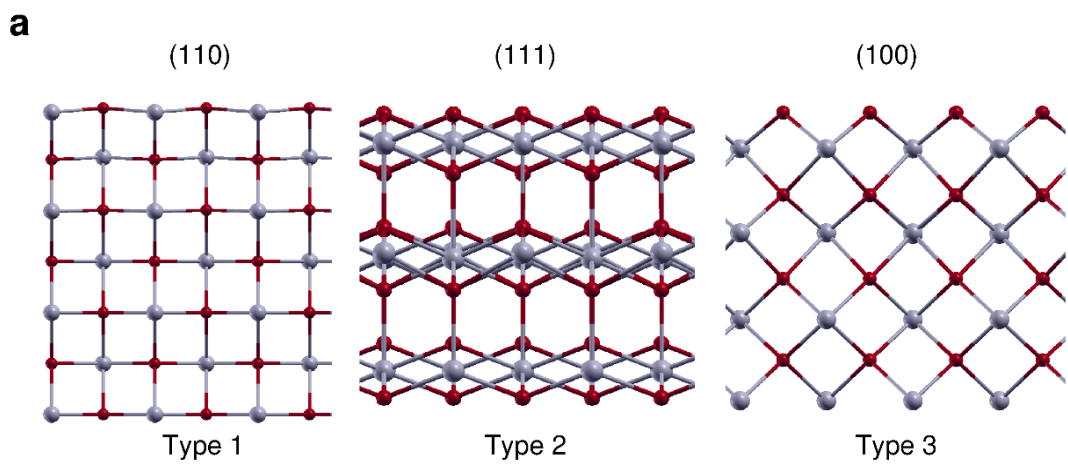

(110)     (111)     (100)

Type 1     Type 2     Type 3

**b**

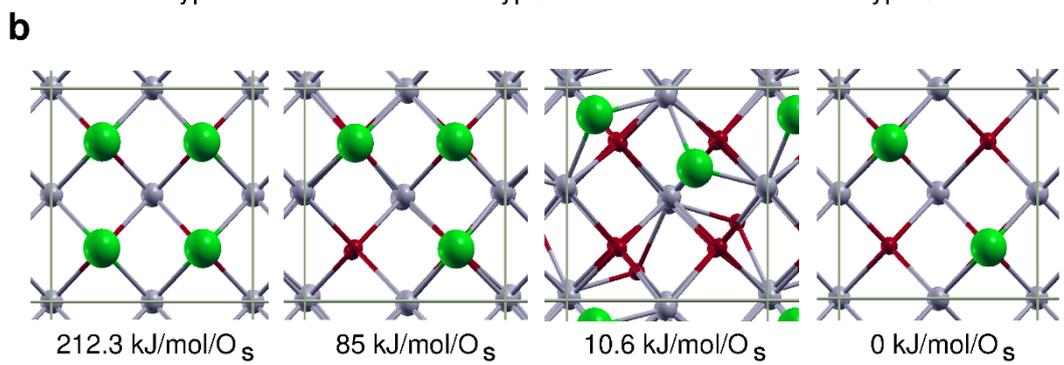

212.3 kJ/mol/$O_s$     85 kJ/mol/$O_s$     10.6 kJ/mol/$O_s$     0 kJ/mol/$O_s$

**c**

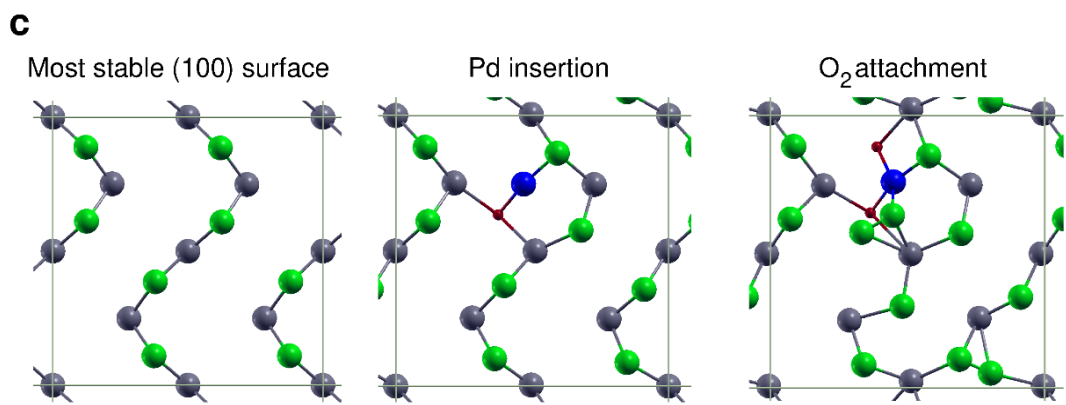

Most stable (100) surface     Pd insertion     $O_2$ attachment

**d**

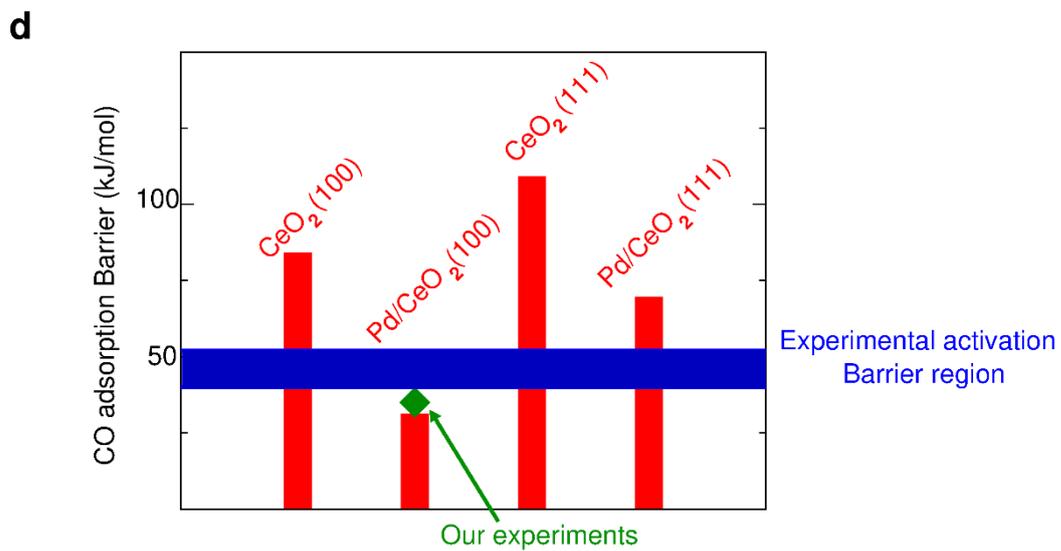

Figure 1. Structure of Pd/CeO$_2$ and the CO adsorption barriers. (a) Tasker's stacking sequence types and low-index CeO$_2$ (110), (111), (100) surfaces. Red and gray spheres represent oxygen and cerium atoms, respectively. (b) Relative energy compared to the lowest energy configuration on a 1×1 (100) surface; green spheres represent the topmost surface oxygen atoms; O$_s$ represents surface oxygen. (c) The most stable CeO$_2$ (100) 2×2 surface structure (left panel) and its modification by Pd insertion (central panel), and molecular oxygen attachment on the Pd atom (right panel) are shown according to the suggested reaction process. Cerium and palladium atoms on the topmost surface are indicated by dark gray and blue spheres, respectively. (d) Calculated CO adsorption barrier (red vertical bars) for each surface structure; the horizontal blue bar indicates the range of the previously reported CO activation barriers and the green diamond shows our experimental barrier.

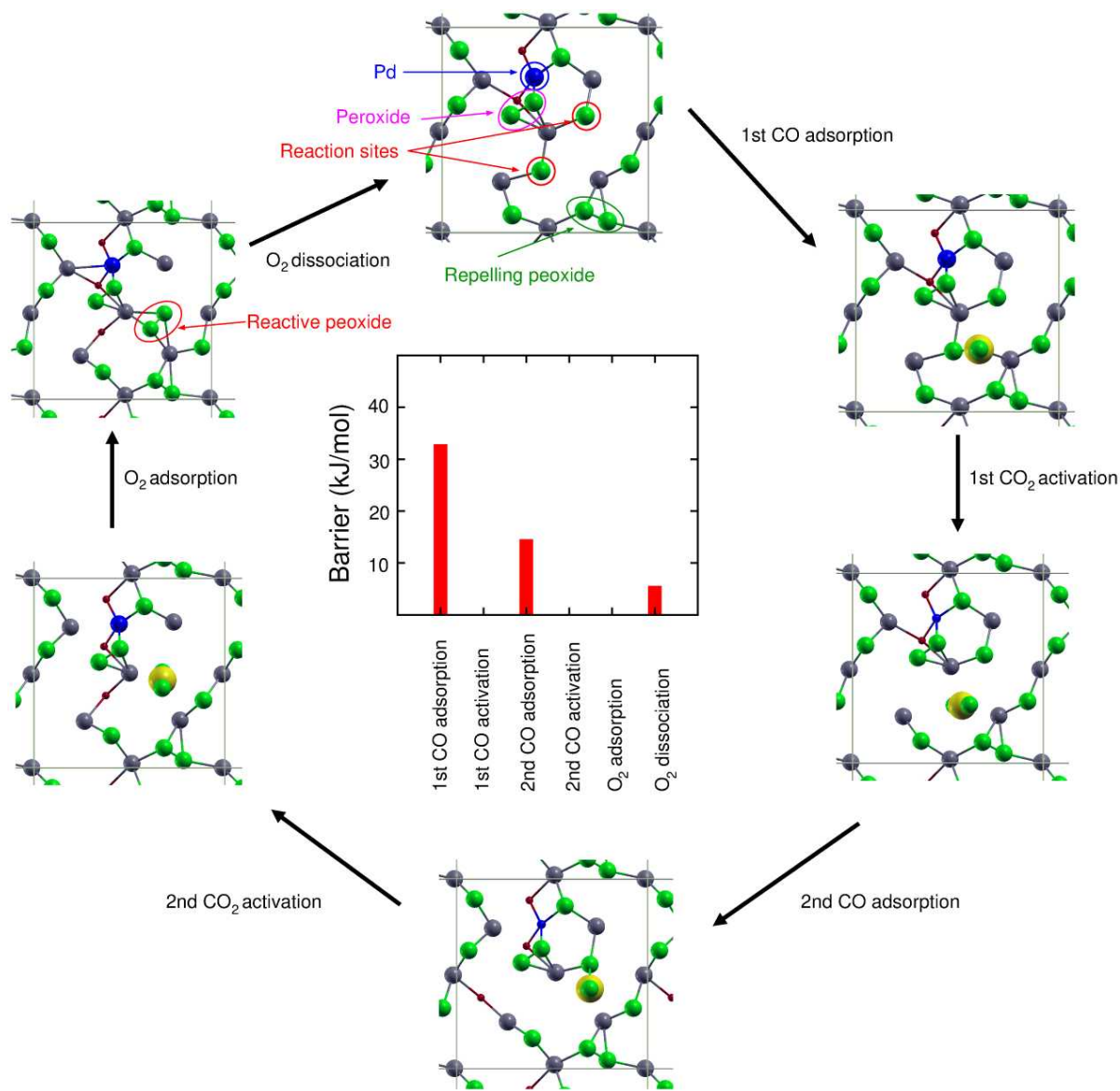

Figure 2. Mechanism of low-temperature CO oxidation on a Pd/CeO$_2$ (100) surface. The initial state is shown at the top, and the arrows indicate the reaction flow. The bar diagram in the center shows the reaction barriers corresponding to each step. Blue, green, dark gray, and yellow spheres represent Pd, surface oxygen, surface Ce, and carbon atoms, respectively. Subsurface O atoms are represented by red spheres.

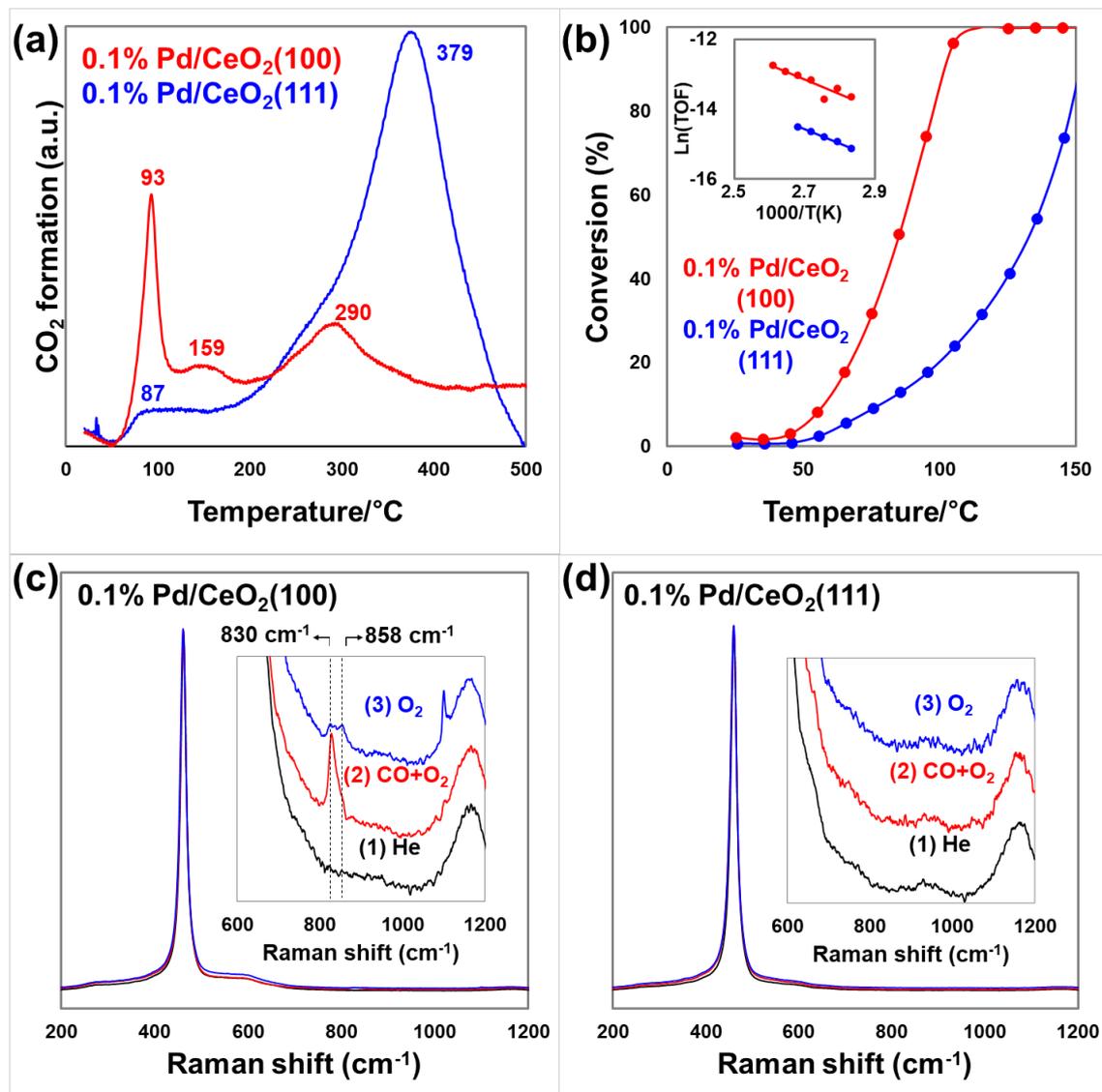

Figure 3. Experimental results for 0.1% Pd/CeO$_2$ catalysts. Profiles for (a) CO-temperature-programmed reduction and (b) temperature-programmed CO oxidation (inset: Arrhenius plots for CO oxidation). (c), (d) *In-situ* Raman spectra of 0.1% Pd/CeO$_2$ catalysts (inset: magnified views).